\documentclass[a4paper, 11pt]{article}
\usepackage{amsmath,amssymb,comment}
  \usepackage{paralist}
  \usepackage{graphics} 
  \usepackage{epsfig} 
\usepackage{graphicx}  \usepackage{epstopdf}
 \usepackage[colorlinks=true]{hyperref}
\hypersetup{urlcolor=blue, citecolor=red}

\newcommand{\Beq}{\begin{equation}}
\newcommand{\Eeq}{\end{equation}}
\newcommand{\BS}{\begin{subequations}}
\newcommand{\ES}{\end{subequations}}
\newcommand{\Beqn}{\begin{equation*}}
\newcommand{\Eeqn}{\end{equation*}}
\newcommand{\Beqa}{\begin{eqnarray}}
\newcommand{\Eeqa}{\end{eqnarray}}
\newcommand{\Beqan}{\begin{eqnarray*}}
\newcommand{\Eeqan}{\end{eqnarray*}}

\begin{document}
\title{One-dimensional weakly nonlinear model equations for Rossby waves}




\maketitle

\medskip

\centerline{\scshape David Henry}
\medskip
{\footnotesize
 \centerline{School of Mathematical Sciences}
   \centerline{University College Cork}
   \centerline{Cork, Ireland}
    \centerline{d.henry@ucc.ie}
}
\medskip
\centerline{\scshape Rossen Ivanov}
\medskip
{\footnotesize
 \centerline{ School of Mathematical Sciences}
\centerline{Dublin Institute of Technology,  Kevin Street}
   \centerline{Dublin 8, Ireland}
   \centerline{rossen.ivanov@dit.ie}
}
\medskip

2010 MSC, Primary: 35Q35, 35Q51, 35Q53; Secondary: 37K10.

\begin{abstract}
In this study we explore several possibilities for modelling
weakly nonlinear Rossby waves in fluid of constant depth, which
propagate predominantly in one direction. The model equations
obtained include the BBM equation, as well as the integrable KdV
and Degasperis-Procesi equations.
\end{abstract}

\section{Introduction}
In this paper we present (formal) derivations for  one-dimensional, weakly nonlinear model equations, in the geophysical setting of Rossby waves. Our method of derivation, which relies on techniques from asymptotic perturbation theory, leads us to the celebrated KdV and BBM
equations \cite{KdV,BBM72}, or to the Degasperis-Procesi (DP) equation \cite{DP}, depending on the order of terms included in the
model. It is the derivation of the DP equation in this geophysical context which is of particular interest.  We will see during the course of our derivation that the structural properties of the DP equation are compatible, and consistent, with the geophysical problem we consider, in the sense that the dispersion relation for the DP equation matches that of the Rossby wave we examine.

The Earth's oceans and atmospheres, in general, behave as a strongly turbulent media. However, on large spatial, and slow time, scales highly ordered self organised structures can emerge. This can be observed for myriad phenomena in geophysical fluid mechanics \cite{CR,P79,R83}, among them being  Rossby waves. Geophysical fluid dynamics is the study of fluid motions where the Earth's rotation, in particular the Coriolis forces, plays a significant role in the governing equations \cite{CR,P79,R83}. Rossby waves are geophysical waves which plays a role in modelling the oceanic Gulf Stream, zonal winds,  climate variability propagating from the tropics during ENSO events, and equatorial waves, cf. \cite{B80,CR,Fed,P79,R83}.

To the leading order, Rossby waves are governed by linear model equations \cite{CR,P79,R83}, which will be introduced in Section \ref{SecGov}. If weak nonlinearities are incorporated into the governing equation, through  asymptotic analysis, then it is known that nonlinear models which are structurally simililar to the KdV and mKdV equations may be obtained, cf. \cite{B80}. We also note that two-dimensional nonlinear models for Rossby waves have been studied in \cite{G03}.  The KdV and mKdV model equations are integrable, with extremely rich and profound structural properties, and they have solutions in the form of solitary waves. The solitary waves arise as a balance between the nonlinear and dispersive terms in the equation and are stable in time, so they are particularly suitable for the description of emerging structures.

In Section \ref{SecKB} we present an alternative derivation of the KdV equation for Rossby waves, and furthermore we derive the BBM equation at the same order of asymptotic expansion. The BBM equation is a much celebrated model equation in the theory of water waves, with many interesting properties. However, a significant drawback of the BBM equation is that it is not integrable.

In Section \ref{SecDP}  we formally derive the DP equation at the next order of the asymptotic expansion. The DP equation \cite{DP} is a recently derived nonlinear dispersive partial differential equation which has many interesting properties: it is an integrable equation, however unlike the KdV equation it also exhibits wave-breaking. By wave breaking, we mean the phenomenon whereby  the solution
remain bounded, but its slope becomes unbounded, in finite time, cf. \cite{ConEschActa}. It belongs to the so-called $b-$family of equations, of which the Camassa-Holm (CH) equation is the only other integrable equation \cite{I07}, and both of these equations may be derived in the context of gravitational shallow water waves \cite{ConLan,J02}. Both the DP and CH equations share a number of interesting and rich structural properties, including integrability as infinite-dimensional Hamiltonian systems \cite{1,2,3,4,5}, persistence properties of solutions \cite{C5,HenNon,HenDCD,HenDCDi,H5,M}, geometrical interpretations \cite{HMR98,ConKol2003,EK,K}, and, of particular interest for the fluid dynamics community, they exhibit both global solutions and solutions which undergo wave breaking \cite{ConEschActa,ELY}.  A  further aspect of CH and DP that attracted a lot of attention is the existence of peaked travelling waves (peakons), whose shape is stable under small perturbations (orbital stability) cf. \cite{6,7}. Particularly interesting is the fact that the peakons have the same shape as the travelling waves of greatest height that solve the governing equations for water waves - see \cite{8,9,10}.

An interesting aspect of the CH and DP equations as models for shallow water waves is that their dispersion laws differ from the typical one for small-amplitude shallow water waves. This apparent disparity is accounted for by the analysis  in \cite{ConLan}, where it was shown that both CH and DP arise in the regime of shallow water waves of moderate (not small) amplitude. A nice artefact of our formal derivation of the DP equation for Rossby waves, in Section \ref{SecDP}, is that that the dispersion law for the DP equation is consistent with the dispersion relation for Rossby waves. The conclusion is that the geophysical setting is highly suitable for the DP as a model equation, and this paper may be regarded as a first approach in investigating whether the DP, and perhaps the CH, equations play a role as model equations in the geophysical setting.

\section{Preliminaries}\label{SecGov}
\subsection{Governing equations}
In this Section we present a brief derivation of the Rossby wave equation, starting from the full governing equations for geophysical water waves. The full governing equations for geophysical water waves are highly intractable partial differential equations, being highly nonlinear, and with additional complications such as an unknown free-surface. It is worthy of mention, however, that a body of recent work has derived exact solutions for the full governing equations at the equator \cite{11,C13,ConJGR,ConJGL,H,Mat}. The standard approach in working with the governing equation is to somehow simplify the full equations, for example, through assuming  geostrophic balance, or through linearisation. The Rossby wave is an artefact of linearisation of the governing equations, and play a role in modelling a number of geophysical processes. While the Rossby wave may be derived in a number of different fashions as a quasi-geostrophic wave, for instance, it may be derived as a by-product of topographic changes in the sea-bed, we present below a derivation of the Rossby wave over a flat sea-bed at mid-latitudes--- such waves play an interesting role in interactions with the Antarctic Circumpolar Current \cite{OWA06}, for instance.

  We take the earth to be a perfect sphere of radius $R=6378$ $km$, which has a constant rotational speed of $\Omega=73.10^{-6}$ $rad/s$, and $g=9.8$ $ms^{-2}$ is the standard gravitational acceleration at the earth's surface \cite{CR}. From the viewpoint of a rotating reference frame with its origin at the earth's surface,  so that the $\{x,y,z\}$-coordinate frame is chosen with $z$ as the vertical variable, $x$ as the longitudinal variable (in the direction due east), and $y$ is the latitudinal variable (in the direction due north), then the governing equations for geophysical ocean waves are given by \cite{CR}
\begin{subequations}\label{Gov}
\begin{subequations}\label{GovA}
\begin{align}
u_t+ uu_x+vu_y+wu_z +f_{*} w-f v &=-\frac 1 \rho P_x,   \\
v_t+uv_x+vv_y+wv_z+f u&=- \frac 1 \rho P_y,  \\
w_t+uw_x+vw_y+ww_z-f_{*}u&=- \frac 1 \rho P_z-g,
\end{align}
\end{subequations}
together with the equation for mass conservation
\Beq\label{mc}
\rho_t+u\rho_x+v\rho_y+w\rho_z=0
\Eeq
and the equation of incompressibility
\Beq\label{in}
u_x+v_y+w_z=0.
\Eeq
\end{subequations}
Here $(u,v,w)$ is the velocity field of the fluid,  $\rho$ is the density of the fluid, $P$ is the pressure of the fluid, and we have neglected viscous effects in the fluid: we assume the fluid is inviscid and incompressible. The Coriolis parameter $f$, and the so-called reciprocal Coriolis parameter $f_{*}$, are defined by
\Beq \label{f}
f=2\Omega  \sin \phi, \qquad f_{*}=2 \Omega \cos \phi,
\Eeq
where the variable $\phi$ represents the latitude. Geophysical waves, for which the Coriolis effects of the Earth's rotation have a significant impact on the fluid motion, typically occur when $\frac{U}{\lambda}\lesssim \Omega$ \cite{CR}, or expressed in terms of the Rossby number $Ro$,
\[
Ro=\frac{U}{\lambda \Omega }=\mathcal O(1).
\] Here $\lambda$ is the typical horizontal length scale of the flow, and $U$ is the typical horizontal velocity scale for the fluid motion, and we take the symbol $\mathcal O(1)$ to mean that the term is of the order of magnitude of one, or less.

The full governing equations \eqref{Gov} are highly nonlinear and intractable, and it would be very desirable, and indeed necessary, for us to simplify them somehow. Thanks to the high  variance in orders of magnitude of the physical scales which are inherent in geophysical waves, simplification can be achieved in a reasonable fashion for certain wave regimes if we perform an analysis of the comparative orders of magnitude of the physical  scales for each of the variables and terms of the equations \eqref{Gov}. For instance, it can be easily deduced, from an analysis of the terms in \eqref{Gov}, that large (horizontal)  scale geophysical flows must be shallow, in the sense that $H \ll \lambda$, where $H$ is the typical vertical depth of the fluid \cite{CR}. Additionally, large scale geophysical flows are almost purely horizontal, that is $W \ll U$,  where $W$ is the typical vertical velocity scale \cite{CR}. Furthermore, away from the equator (where additional factors must be incorporated into the scale comparison), the terms involving $f_*$ play an insignificant role in the governing equations. Following the scaling analysis, the equations \eqref{GovA} may be reduced to
\begin{subequations}\label{Gov'}
\begin{align}
\label{u1} u_t+ uu_x+vu_y+wu_z -f v &=-\frac 1 \rho p_x,   \\
\label{v1} v_t+uv_x+vv_y+wv_z+f u&=- \frac 1 \rho p_y,
\end{align}
and
\Beq
\label{p} p_z=0,
\Eeq
\end{subequations} where $p=P-P_0-\rho g(H-z)$ is the dynamic pressure.
Now, if $u_z\equiv v_z\equiv 0$ initially, then from \eqref{p}, \eqref{u1} and \eqref{v1} it follows that $u_z\equiv 0,v_z \equiv 0$ for all times, that is, the horizontal flow remains independent of depth: these flows are known as barotropic.
In the following, we let $z=0$ denote the flat sea bed, and take the mean-depth of the fluid as the vertical length scale $H$.
\subsection{Rossby waves}
In certain geophysical regimes, the nonlinear advective terms in the governing equations play a negligible role in the underlying motion. Specifically, this scenario occurs when the Rossby-number is low, $Ro\ll 1$, and the so-called temporal Rossby number $Ro_T=1/\Omega T \sim 1$, where $T$ is the typical time-scale for the flow \cite{CR}.  According, the governing equations can be expressed as
\begin{subequations}\label{Gov''}
\begin{align}
\label{u1'} u_t  -f v &=-g\eta_x,   \\
\label{v1'} v_t+f u&=- g\eta_y,
\end{align}
\end{subequations}
where $\eta(x,y,t)$ represents the surface wave over the flat bed. The equations \eqref{Gov''} govern a number of different geophysical phenonema: Kelvin (trapped) waves, Poincar\'e waves, and Rossby (planetary) waves. Rossby wave solutions of \eqref{Gov''} are waves which exhibit a slow evolution from steady geostrophic flows, and are accordingly termed quasi-geostrophic waves. Geostrophic flows are those where the Coriolis force dominates the acceleration terms in the governing equations, and so the pressure gradient force is balanced by the Coriolis effect. We present a derivation of the governing equations for Rossby waves which occur at mid-latitudes. We note that there are also equatorial Rossby waves  \cite{Fed}, which share some  characteristics with the waves we will present.

At mid-latitudes, the Coriolis parameter $f$ in \eqref{f} can be expanded in terms of a Taylor series for $\phi=\phi_0+y/R$, where $\phi_0$ is a reference latitude and $R$ is the Earth's radius. Retaining the first two terms we have the  $\beta-$plane approximation
\Beq
f=f_0+\beta_0 y,
\Eeq
and if we omit the second term we have the  $f-$plane approximation.
Here $f_0=2\Omega \sin \phi_0$ is the reference Coriolis parameter, $\beta_0=2(\Omega/a)\cos \phi_0$ is the $\beta-$parameter. The $\beta-$plane approximation is valid when the dimensionless planetary number $\beta=\beta_0 \lambda/f_0 \ll 1$, which gives us restrictions on the reference latitudes $\phi_0$ and meridional length scales $L$ for which Rossby waves apply \cite{CR}: typical values for these parameters are $f_0=8\times 10^{-5}s^{-1}$ and $\beta_0=2\times 10^{-11}m^{-1}s^{-1}$, and so obviously $\beta \ll 1$ for all waves of length $\lambda \ll 4000 km$.
Applying a formal asymptotic analysis of the equations \eqref{u1'}-\eqref{v1'}, we get expressions for $u,v$ in terms of partial derivatives of $\eta$. Substituting the resulting expressions into
\Beq
\label{cont'} \eta_t+H(u_x+v_y)=0
\Eeq gives us the following linearised Rossby wave equation for the free surface height $\eta$:
\begin{equation}\label{RossbyLinear}
\eta_t-L_D^2\Delta_{x,y}\eta_t-\beta L^2_D\eta_x=0.
\end{equation}
Here $L_D=\sqrt{gH}/f_0$ is the Rossby deformation radius. The
waveform  $$\eta\sim e^{i(\omega(k,n)t-kx-ny)}$$ is a solution of
\eqref{RossbyLinear} if the dispersion relation is satisfied:
\begin{equation}\label{omega}
\omega(k,n)=-\frac{\beta_0 k}{k^2+n^2+L_D^{-2}}.
\end{equation}

\section{Weakly nonlinear models: BBM and KdV}\label{SecKB}
In order to model  zonal Rossby wave flows, which are largely one-dimensional, we derive obtain a suitable model equation from \eqref{RossbyLinear} using the ansatz  $\tilde\eta =\cos (ny)\, {\eta}(x,t)$ for  waves in a channel; $\eta(x,t)$ then satisfies the equation
\begin{equation}\label{RossbyLinear2}
\left(1+L_D^2(n^2-\partial_x^2)\right)\eta_t-\beta_0 L_D^2 \eta_x=0.
\end{equation}
The one-dimensionality of the motion is a common characteristic of long waves, and it is relevant if one considers simply $x$-directional
propagation at fixed latitude, when the wavelength
$\lambda>>L_D$.
With $n$ fixed, in this one-dimensional
model  $\omega=\omega(k)$ depends only on the $x$-directional wave number $k$. We note that, for long waves, $k\sim 2\pi/\lambda\ll 1$.
In general, one dimensional model equations with nonlinearity and
dispersion have the universal form \cite{W67,W74}:
\begin{equation}\label{general}
 \begin{split}
u_t + uu_x + \kappa \int_{-\infty}^{\infty}K(x-x')u_x(x',t)dx'=&0,
\\
K(x)=\int_{-\infty}^{\infty} e^{ikx}c(k) \frac{dk}{2\pi}, &
 \end{split}
\end{equation}
where $ \omega(k)=kc(k),$ is the dispersion law of the model. For
example, $\omega(k)\sim k^3 $ for KdV equation, $\omega(k)\sim
k|k|$ for the Benjamin-Ono equation, etc.

The leading order terms in equation (\ref{RossbyLinear2}) are
\begin{equation}\label{RossbyLinear3} (1+L_D^2 n^2)\eta_t-\beta_0 L_D^2\eta_x\sim
0,
\end{equation} with $\eta_{xxt}$ of smaller order.

Furthermore, there is a nonlinear term present, $\epsilon \eta
\eta_x $, at the leading approximation (and yet small) as it
transpires from (\ref{general}). With this term the weakly
nonlinear equation becomes
\begin{equation}\label{RossbyNonLinear2} \eta_t-\beta_0 L_D^2
\left(1+L_D^2(n^2-\partial_x^2)\right)^{-1}\eta_x+
 \epsilon \eta \eta_x=0,
\end{equation} where $\epsilon=\epsilon(n)$ is a dimensional constant for fixed $n$.
Then \begin{equation}\label{RossbyNonLinear3}
\left(1+L_D^2(n^2-\partial_x^2)\right)\eta_t-\beta_0 L_D^2 \eta_x+
 \epsilon\left(1+L_D^2(n^2-\partial_x^2)\right) \eta \eta_x=0,
\end{equation} or \begin{equation}\label{RossbyNonLinear4}
\left(1+L_D^2(n^2-\partial_x^2)\right)\eta_t-\beta_0 L_D^2\eta_x+
\epsilon\left(1+L_D^2 n^2\right)\eta \eta_x- \epsilon L_D^2
(3\eta_x\eta_{xx}+\eta\eta_{xxx}) =0,
\end{equation} Neglecting smaller order $\epsilon L_D^2$ terms we can write the
model equation as \begin{equation}\label{BBM} (1+L_D^2
n^2)\eta_t-\beta_0 L_D^2\eta_x -L_D^2\eta_{xxt}+
\epsilon\left(1+L_D^2 n^2\right)\eta \eta_x=0,
\end{equation} which is known as the BBM equation \cite{12,BBM72}. Indeed, if one
changes  for simplicity $x\to x+\frac{\beta_0 L_D^2}{1+L_D^2 n^2}
t$, the equation acquires its recognizable form
\begin{equation}\label{BBM2} (1+L_D^2 n^2)\eta_t -L_D^2\eta_{xxt}+
\epsilon\left(1+L_D^2 n^2\right)\eta \eta_x=0.
\end{equation}
A major benefit of the model equation (\ref{BBM}) is that its dispersion law
matches exactly the one for Rossby waves (\ref{omega}). The BBM equation is not integrable, yet it allows
soliton-like solutions and is very convenient for numerical simulations.
It can be
transformed to the integrable KdV equation, keeping terms to a
certain order, as follows.
The term $\eta_{xxt}$ is of smaller order, however of the same order as $\eta_{xxx}$. Using Johnson's approach \cite{J02} the two terms can be replaced with each other in the model equation. More rigorously, estimates about the difference between the two terms are studied in \cite{BPD83}. Then we can use (\ref{RossbyLinear3}) in (\ref{BBM2}) and for an
arbitrary $\mu$ we can write \begin{equation}\label{BBM3} (1+L_D^2 n^2)\eta_t
-(L_D^2+\mu)\eta_{xxt}+\mu \frac{\beta_0 L_D^2 }{1+L_D^2
n^2}\eta_{xxx} +\epsilon\left(1+L_D^2 n^2\right)\eta \eta_x=0.
\end{equation} The choice $\mu=-L_D^2$ removes the $\eta_{xxt}$ term and we
obtain the KdV equation \cite{KdV}
\begin{equation}\label{KdV}
(1+L_D^2 n^2)\eta_t - \frac{\beta_0 L_D^4 }{1+L_D^2 n^2}\eta_{xxx}
+\epsilon\left(1+L_D^2 n^2\right)\eta \eta_x=0,
\end{equation}
\noindent or, with the linear dispersion term,
\begin{equation}\label{KdV2}
(1+L_D^2 n^2)\eta_t -\beta_0 L_D^2\eta_x - \frac{\beta_0 L_D^4
}{1+L_D^2 n^2} \eta_{xxx} +\epsilon\left(1+L_D^2 n^2\right)\eta
\eta_x=0,
\end{equation}
The KdV is an integrable equation \cite{GGKM,AC11,J97,ZMNP},
however its dispersion relation $\omega(k)\sim k^3 $ (or, if $\eta_x$
term is included, $\omega(k)\sim k+ \nu k^3 $, $\nu$ being a
constant) differs from (\ref{omega}) and conforms with
(\ref{omega}) only for small $k$. Of course this is consistent
with the long waves assumption, that $\lambda\sim \frac{2\pi}{k}$
is large. Another approach for modelling Rossby waves which leads to the KdV
equation is given in \cite{B80}. The solution there also has the
form of a product of two functions, one of which depends on $y$
and the other satisfying the KdV equation with coefficients,
depending on $n$.

\section{Other integrable models: The Degasperis-Procesi equation}\label{SecDP}

In this Section we present a formal derivation of the DP equation
from the Rossby wave equation \eqref{RossbyLinear2}. Applying
again Johnson's trick on the term $\eta_{xxt}$ of
(\ref{RossbyNonLinear4}) for an arbitrary $\mu$, and taking into
account (\ref{RossbyLinear3}) we obtain
\begin{equation}\label{RossbyNonLinear5}
\begin{split}
\left(1+L_D^2n^2\right)\eta_t-&(L_D^2+\mu)\eta_{xxt}+\frac{\mu
\beta_0
L_D^2 }{1+L_D^2n^2}\eta_{xxx}-\beta_0 L_D^2\eta_x\\
+&\epsilon\left(1+L_D^2 n^2\right)\eta \eta_x- \epsilon L_D^2
(3\eta_x\eta_{xx}+\eta\eta_{xxx}) =0,
\end{split}
\end{equation}
The $\eta_{xxx}$ term can be removed by a constant shift
transformation,
$$\eta \to \eta+\frac{\mu \beta_0}{\epsilon(1+L_D^2n^2)},$$ leaving
\begin{equation}\label{RossbyNonLinear6}
 \begin{split}
\left(1+L_D^2n^2\right)\eta_t & -(L_D^2+\mu)\eta_{xxt}+\beta_0(\mu- L_D^2)\eta_x \\
&+\epsilon\left(1+L_D^2 n^2\right)\eta \eta_x- \epsilon L_D^2
(3\eta_x\eta_{xx}+\eta\eta_{xxx}) =0,
\end{split}
\end{equation}
Rescaling $$ \partial_x \to \sqrt{\frac{1+L_D^2n^2}{L_D^2+\mu}}
\partial_x \qquad \partial_t\to \sqrt{\frac{1+L_D^2n^2}{L_D^2+\mu}}
\partial_t $$ gives
\begin{equation}\label{RossbyNonLinear7}
\left(1-\partial_x^2\right)\eta_t+\beta_0\frac{\mu-L_D^2}{1+L_D^2n^2}\eta_x+
\epsilon\eta \eta_x-  \frac{\epsilon L_D^2}{L_D^2+\mu}
(3\eta_x\eta_{xx}+\eta\eta_{xxx}) =0,
\end{equation}
We can choose $\mu=3L_D^2$, then
\begin{equation}\label{RossbyNonLinear8}
\left(1-\partial_x^2\right)\eta_t+\beta_0\frac{2L_D^2}{1+L_D^2n^2}\eta_x+
\epsilon\eta \eta_x- \frac{\epsilon }{4}
(3\eta_x\eta_{xx}+\eta\eta_{xxx}) =0,
\end{equation}
The last equation up to rescaling $\eta \to 4\eta / \epsilon $ is
the Degasperis-Procesi (DP) equation \cite{DP}, since the ratio of
the coefficients in the last bracket is 3:1.
\begin{equation}\label{RossbyNonLinear9}
\left(1-\partial_x^2\right)\eta_t+\beta_0\frac{2L_D^2}{1+L_D^2n^2}\eta_x+
4\eta \eta_x- (3\eta_x\eta_{xx}+\eta\eta_{xxx}) =0,
\end{equation} which can be written also as \begin{equation}
\label{DPeq} m_t + \kappa \eta_x+3m\eta_x + \eta m_x = 0, \qquad m
= \eta -\eta_{xx},
\end{equation} with $\kappa=\beta_0\frac{2L_D^2}{1+L_D^2n^2}$ being
a constant.

The DP is an integrable equation for any constant $\kappa$,
including $\kappa=0$ \cite{CIL,DHH,DP,I07}, in the dispersionless case
$\kappa=0$ it allows peakon solutions, otherwise it has also
smooth soliton solutions as well as breaking waves \cite{ELY}, subject to
special initial conditions.
The DP equation together with the Camasa-Holm (CH) equation
\cite{CH,J02,I07,HI11,AC11}, \begin{equation} \label{CHeq}
\eta_t-\eta_{xxt}+\kappa \eta_x + 3\eta \eta_x -2\eta_x\eta_{xx} -\eta
\eta_{xxx} = 0
\end{equation} ($\kappa$ is an arbitrary constant related to the linear
dispersion term) which has a similar nonlinear structure, have
been put forward as shallow water models \cite{ConLan,J02,I07}.
Both equations are integrable and share some common features, such
as breaking waves \cite{ConEschActa,ELY}, geometrical
interpretations \cite{ConKol2003,EK,K} and persistence properties
of solutions \cite{C5,HenNon,HenDCD,HenDCDi,H5,M}. While their
applicability as shallow water models has been well-studied, it was shown in \cite{ConLan} that both CH and DP arise in the regime of shallow water waves of moderate (not small) amplitude. This explains the disparity of the dispersion laws for the CH and DP equations, and the typical one for small-amplitude shallow water waves, given by $$
\omega(k)=\sqrt{gk\tanh(kh)}.$$ In this study we argue that the DP
equation is well suited to the geophysical setting of Rossby waves, since the dispersion
relation of the Rossby waves (\ref{omega}) is in the same form as
for the DP equation. Indeed, if one neglects the nonlinear terms
in the equations (\ref{DPeq}), (\ref{CHeq}) their linearizations
admit wave solutions of the form $\eta\sim \exp(i\omega(k) t -
ikx)$, such that \begin{equation} \nonumber \omega(k)=\frac{\kappa
k}{1+k^2}
\end{equation} which up to rescaling of the constants gives the dispersion law
(\ref{omega}) in the case when $n$ is constant.

\section{Conclusions} We have derived several model equations for
weakly nonlinear Rossby waves propagating predominantly in one
direction. Two approximate models including same order of terms
relate to the BBM equation and to the integrable KdV equation with
coefficients, depending on the wave number $n$ of the transversal
direction $y$. Both BBM and KdV equations have certain advantages
from analytical and computational point of view. Another
previously known similar model of Rossby waves \cite{B80} in
slightly different set up presents the solution in a form of a
product of two functions, one of which depends on $y$ and the
other satisfying the KdV or mKdV equations with coefficients,
depending on $n$.

When terms of higher order of smallness are kept, the model can be
transformed to the integrable DP equation which has been
extensively studied recently both analytically and in the context
of water waves. The usefulness of the DP equation in the modelling
of Rossby waves is due to its dispersion law, which has the same
form as the dispersion relation of the Rossby waves (\ref{omega}).
We can speculate that the integrable Camassa-Holm equation, which has the same dispersion law as the DP equation and the same type of nonlinearities can also be put forward as a model for Rossby waves. To this end a rearrangement of the nonlinear terms of the smallest order might be necessary following the techniques from the work of Johnson \cite{J02}. In comparison to KdV equation, CH and DP equations have more interesting and involved features, such as allowing for both stable and wave-breaking solutions. Indeed, the CH and DP are  archetypes for nonlinear dispersive partial differential equations which are both integrable (and so possess the associated rich structure) and allow for wave-breaking.


\section*{Acknowledgments} We would like to thank Prof. A.
Constantin for bringing to our attention the problem treated in
this paper and for some very helpful discussions. We are thankful
to the two anonymous referees for their valuable comments and
suggestions. The second author is supported by SFI grant
09/RFP/MTH2144.


\end{document}